\documentclass[runningheads]{llncs}
\usepackage{graphicx}
\usepackage{tabularx}
\usepackage{lipsum}
\usepackage{graphicx}
\usepackage{amssymb}
\usepackage{amsmath}
\usepackage{cite}
\usepackage{epstopdf}
\usepackage{mathrsfs}
\usepackage{mathtools}
\usepackage[justification=centering]{caption}
\usepackage{stackengine}
\usepackage{cite}
\usepackage{color}
\usepackage{url}
\begin{document}
\title{Secrecy Outage Probability of Cognitive Small-Cell Network with Unreliable Backhaul Connections\thanks{This work was supported in part by the Royal Society-SERB Newton International Fellowship under Grant NF151345}}
\author{Jinghua Zhang \and
Chinmoy Kundu \and
Emi Garcia-Palacios}
\authorrunning{Jinghua Zhang}
\institute{Queen's University Belfast, Belfast, United Kingdom, BT9 5AH\\
\email{jzhang22@qub.ac.uk, c.kundu@qub.ac.uk, e.garcia@ee.qub.ac.uk.} }
\maketitle              
\begin{abstract}
In this paper, we investigate the secrecy performance of underlay cognitive small-cell radio network with unreliable backhaul connections. The secondary cognitive small-cell transmitters are connected to macro base station by wireless backhaul links. The small-cell network is sharing the same spectrum with the primary network ensuring that a desired outage probability constraint in the primary network is always satisfied. We propose an optimal transmitter selection (OTS) scheme for small-cell network to transfer information to the destination. The closed-form expression of secrecy outage probability are derived. Our result shows that increasing the primary transmitter's transmit power and the number of small-cell transmitter can improve the system performance. The backhaul reliability of secondary and the desired outage probability of the primary also have significant impact on the system.
\end{abstract}
\section{Introduction}
Due to the explosion of data-intensive applications and wireless systems such as the Internet of Things (IoT) and smart cities, the deployment of wireless infrastructure is expected to get more dense and heterogeneous in the near future \cite{Jeffrey20145G}. To reach such high data rate, the backhaul links connecting the macro-cell and many small-cells in the heterogeneous networks (HetNets) are also expected to become dense. In the conventional wired backhaul network, high reliability wired links and high data rate can be expected, however, the deploying and sustaining the large-scale wired links requir excessive capital investment for all the connections \cite{Orawan2011Evolution,Xiaohu20145G}. This leads wireless backhaul as alternative solution since it has been proven cost-effective and flexible in practical systems. However, wireless backhaul is unlikely reliable as wired backhaul due to non-line-of-sight (n-LOS) propagation and fading of wireless channels \cite{Kim2016Secrecy}. 

The aforementioned rapid development in wireless devices and services is also pushing the demand for spectrum while most of licensed spectrum bands are occupied \cite{Kolodzy2002spectrum}. In recent years, the investigation on cognitive radio (CR) techniques \cite{Mitola1999Cognitive} has attracted many experts' attention. CR optimises the current spectrum usage, which allows unlicensed secondary users to share the same spectrum with the licensed primary users in an opportunistic manner. The authors in \cite{Zhangjinghua} analysed the impact of the primary network on the secondary network. In \cite{Lee2015Cognitive}, the authors  
optimizing the time and power allocation in the secondary network. To improve the CR or noncognitive network performance, user selection is always among the secondary users and relays in the literature \cite{Phong2016Secure,Trung2013Cognitive,
Kundu2016Relay,jinghuazhang2017Cognitive}.

For a complete study, we also consider the challenges of security in the wireless communication network. Due to the broadcasting nature of wireless channels, the confidential information in wireless network is vulnerable to eavesdropping and security attacks. In reality, CR networks are easily susceptable to eavesdropping. The conventional way from upper layer security is deploying data encryption for secure communication, on the other hand, physical layer security (PLS) obtain the advantage from the randomness of the wireless channels for information security extensively. PLS has become increasingly popular to deal with wiretapping and possible loss of confidentiality. Some research has investigated the secrecy performance using PLS \cite{Phong2016Secure,Yuzhen2016Secure, Yincheng, Vu2017Secure, Kundu2017AFrelay}. 

Nevertheless, all the aforementioned work did not take into account the impact of unreliable backhaul on PLS of CR network. Some literatures in CR network only consider the interference on the primary network. In some literature on backhaul CR networks \cite{Kim2016Secrecy,Kim2015Performance,
Huy2017Multiuser,Huy2017Secure}, authors investigated the secrecy performance but not consider the system with secondary user selection schemes. The impact of guaranteeing outage as a quality-of-service (QoS) in the primary networks also not considered in the aforementioned paper. Our research address these key issues in CR network with backhaul. We investigate the secrecy performance of CR network with unreliable backhaul connections. Based on those considerations, our contribution of this paper is summarise as follows:
\begin{enumerate}
\item We take into account the backhaul unreliability in secrecy performance. We develop the close-form expression of the secrecy outage probability.
\item We consider interference both in primary and secondary receiver. 
\item We consider primary QoS constraint metric as outage probability, which is different from other works in CR.
\item Our model investigates a small-cell transmitter selection schemes, namely, optimal transmitter selection (OTS) which prioritizes the maximum channel gain S--D, and also assesses the influences of varying the number of small-cell transmitter. 
\end{enumerate} 
The rest of the paper is organised as follow. In section \ref{section 2}, the system can channel models are described. Section \ref{section 3} demonstrates the secrecy outage probability of propose system. Numerical results from monte-carlo simulations are showcased in section \ref{section 5}. Finally, the paper is concluded in section \ref{section 6}. 
\section{System and channel models}\label{section 2}
As illustrated in Fig. 1, the system is consisting of a primary network with one primary transmitter, $\text{T}$, one primary receiver, $\text{R}$ and a secondary network consisting of $K$ small-cells transmitters, $\lbrace \text{S}_1,...,\text{S}_k,...,\text{S}_K\rbrace$ which are connected to a macro-base station, BS, by unreliable backhaul links, one secondary destination, D, and one eavesdropper, E. All nodes are equipped with single antenna. We assume all nodes are sufficiently separated from each other so that $\text{T}-\text{R}$, $\text{T}- \text{D}$, $\text{T}- \text{E}$, $\text{S}-\text{R}$, $\text{S}-\text{D}$ and $\text{S}-\text{E}$ experience independent and identically distributed Rayleigh fading. The channel between nodes are denoted by $h_X$ where X= $\lbrace$ $TR$, $TD$, $TE$, $SR$, $SD$, $SE$ $\rbrace$ channel power gains are exponential distributed with parameters $\lambda_x$ for x=$\lbrace$ $tr$, $td$, $te$, $sr$, $sd$, $se$ $\rbrace$, respectively. The noise at R , E and D is modelled as the additive white Gaussian noise (AWGN) with zero mean and variances $N_0$. One best transmitter will be selected among K small-cell transmitters, to transfer information to D. While the message is sent from the BS to small-cell transmitters,  the backhaul link might have certain probability of failure. Backhaul reliability is modelled as Bernoulli process with success probability, $\mathbb{P}(\mathbb{I}_k=1)=\Lambda$, and failure probability is $\mathbb{P}(\mathbb{I}_k=0)=1-\Lambda$ for each $k=1,\cdots, K$. We investigate the secrecy outage probability (SOP) of the secondary network. 
\begin{figure}
\centering 
\includegraphics[width=4in]{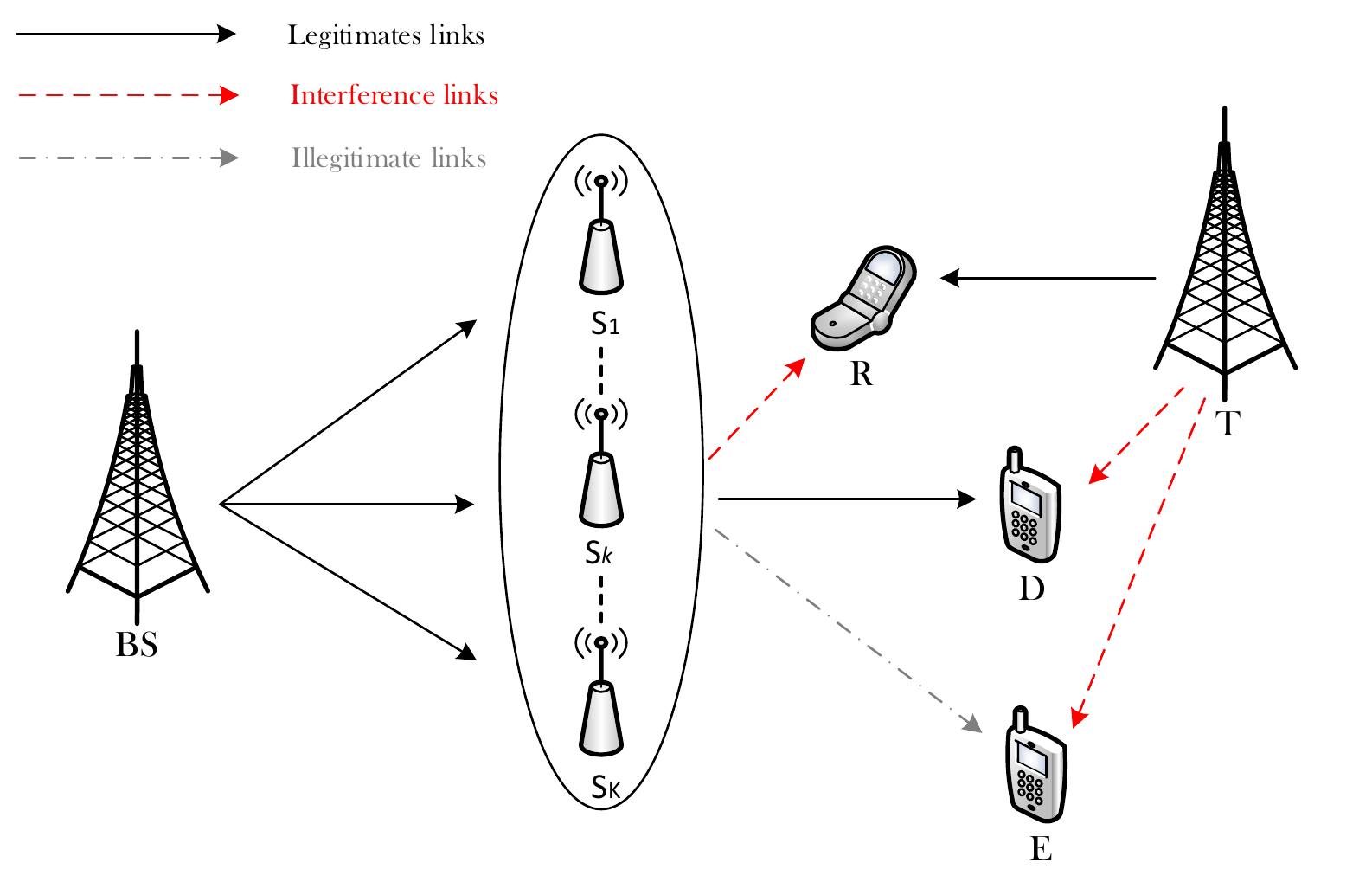} 
\caption{Underlay cognitive radio network with unreliable backhaul connections.}
\label{fig:SM}
\end{figure}
\subsection{Interference at Primary and Secondary Power Constraint}
The primary network is interfered from the selected secondary transmitter, $S_k$, for $k=\lbrace1,...,K\rbrace$ via interference channels $h_{S_kR}$ during the secondary network transmission. The signal-to-interference-plus-noise-ratio (SINR) at $\text{R}$ is given as 
\begin{align}\label{SINR at P_R}
\Gamma_{R}= \frac{P_T|h_{TR}|^2}{P_S|h_{S_kR}|^2+N_0},
\end{align}
where $P_S$ is the maximum allowed transmit power of small-cell transmitter which satisfies the primary network QoS constraint, $h_{TR}$ is the channel coefficient of the $\text{T}- \text{R}$ link, and $h_{S_kR}$ is the channel coefficient of the $\text{S} - \text{R}$ link. 
To protect the primary network, the secondary network transmitters must adapt their transmit power. Moreover, the secondary network transmit power must be limitedly the QoS of the primary network which characterized by its desired outage probability. The primary network outage probability should be below a desired level, $\Phi$. The desired outage probability constraint is defined as follows
\begin{align}\label{desired outage probability}
\mathbb{P}\left[{\Gamma_{\text{R}}}  < \Gamma_0\right]\leq \Phi,
\end{align}
where $\Gamma_0 = 2^{\beta}-1$, $\beta$ is the target rate of the primary network, and $0 < \Phi <1$.
From (\ref{SINR at P_R}) and (\ref{desired outage probability}), output power of secondary transmitter can be derived from the desired outage probability at the primary network
\begin{align}\label{P_H}
    P_S=\left\{
                \begin{array}{ll}
                  P_T\lambda_{sr}\xi,\ \ \ \textnormal{if}\ \ \ \xi>0\\
                   0,\ \ \ \ \ \ \ \ \ \ \ \ \textnormal{otherwise}.\\
                \end{array}
              \right.
\end{align} 
where
\begin{align} \label{xi}
\xi=\frac{1}{\lambda_{tr}\Gamma_0}\left[\frac{\textnormal{exp}\left(\frac{-\lambda_{tr}\Gamma_0}{\Gamma_{T}}\right)}{1-\Phi}-1\right].
\end{align}
Here, we used CDF of $\Gamma_{\text{R}}(x)$ to find $P_S$ in close-form. The CDF of $\Gamma_{\text{R}}(x)$ can be derived from the definition of CDF as
\begin{align} \label{CDF of Psi}
&F_{{\text{R}}}(x)=
1-\frac{\frac{\lambda_{sr}\Gamma_T}{\lambda_{tr}\Gamma_{S}}}{x+\frac{\lambda_{sr}\Gamma_T}{\lambda_{tr}\Gamma_{S}}}\textnormal{exp}\left(\frac{-\lambda_{tr}x}{\Gamma_T}\right),
\end{align}
where $\Gamma_{T}=\frac{P_T}{N_0}$ and $\Gamma_{S}=\frac{P_S}{N_0}$. 
\subsection{Proposed Source Selection and Interference at the Secondary}
To mitigate the eavesdropping, OTS scheme is proposed  where a source is selected to forward the message such that it maximize $\text{S}_k-\text{D}$ link power gain as 
\begin{align}\label{OPS_k}
k^* = \arg \max_{1 \le k \le K} P_S|h_{S_kD}|^2.
\end{align}
Due to the unreliability of the backhaul, the selected link  may not be active.   To consider backhaul reliability into the performance analysis, we model backhaul reliability using Bernoulli random variable $\mathbb{I}$. 
The SINR at D can be given as
\begin{align}
\label{eq_gammasd}
&\Gamma_{SD}=\mathbb{I}{\tilde\Gamma}_{SD},
\end{align}
where \begin{align}
{\tilde\Gamma}_{SD}=\frac{ P_S\max[|h_{S_kD}|^2]}{P_T|h_{TD}|^2+N_0}.
\end{align}
SINR at E can be similarly expressed as
\begin{align}\label{SNR_E}
&\Gamma_{SE}=\dfrac{P_S  |h_{S_{k}E}|^2}{P_T  |h_{TE}|^2+N_0}.
\end{align}
Conditioned on the source has already been selected, E always experience its intercepted signal power as independent exponentially distributed, hence, while finding the distribution of $\Gamma_{SE}$ no backhaul reliability parameter comes into play in (\ref{SNR_E}). However, that is not true for $\Gamma_{SD}$ in (\ref{eq_gammasd}). The distribution of  $\Gamma_{SD}$ will be the mixture distribution of $\mathbb{I}$ and ${\tilde\Gamma}_{SD}$.
Now the distribution of $\Gamma_{SD}$ can be obtained from the mixture distribution due to backhaul reliability as
\begin{align}
\label{eq_delta}
f_{SD}(x)=(1-\Lambda)\delta(x) + \Lambda {\tilde f}_{SD}, 
\end{align}
where $f_{SD}(x)$, ${\tilde f}_{SD}(x)$ are the PDFs of $\Gamma_{SD}$ and  ${\tilde \Gamma}_{SD}$, respectively and $\delta(x)$ is delta function. CDF of $f_{SD}(x)$ can be obtained just by integrating it and finding the CDF of ${\tilde\Gamma}_{SD}$.

The CDF of ${\tilde\Gamma}_{SD}$ can be evaluate from the definition of CDF with the help of the CDF of $\max|h_{S_kD}|^2$ and the PDF of $|h_{TD}|^2$ as
\begin{align}\label{OTS_CDF_SNR_TD}
{\tilde F}_{{SD}}(x)&=\mathbb{P}\left[\frac{ P_S \max|h_{S_kD}|^2 }{P_T|h_{TD}|^2+N_0}<x\right]
\nonumber\\
&=\mathbb{P}\left[ P_S\max_{k=1,...,K}|h_{S_kD}|^2<(P_{T}|h_{TD}|^2+N_0)x\right]
\nonumber\\
&=1-\sum_{k=1}^{K}\binom{K}{k}\frac{(-1)^{k+1}
\frac{\lambda_{td}\Gamma_{S}}{k\lambda_{sd}\Gamma_{T}}}{x+
\frac{\lambda_{td}\Gamma_{S}}{k\lambda_{sd}\Gamma_{T}}}\exp\left( \frac{-k\lambda_{sd}x}{\Gamma_{S}}\right).
\end{align}
The CDF of ${\Gamma}_{SD}$ then can be evaluate with the help of (\ref{eq_delta}) as 
\begin{align}
{F}_{SD}(x)=1-\Lambda \sum_{k=1}^{K}\binom{K}{k}\frac{(-1)^{k+1}
\frac{\lambda_{td}\Gamma_{S}}{k\lambda_{sd}\Gamma_{T}}}{x+
\frac{\lambda_{td}\Gamma_{S}}{k\lambda_{sd}\Gamma_{T}}}\exp\left( \frac{-k\lambda_{sd}x}{\Gamma_{S}}\right).
\end{align}
The CDF of $\Gamma_{S_{k}E}$ can be obtained from the definition of CDF similar to ${\tilde F}_{SD}$ as
\begin{align}\label{CDF_SNR_E}
&F_{{SE}}(x)=1-\frac{\frac{\lambda_{te}\Gamma_{S}}{\lambda_{se}\Gamma_{T}}}{x+\frac{\lambda_{te}\Gamma_{S}}{\lambda_{se}\Gamma_{T}}} \exp\left(\frac{-\lambda_{se}x}{\Gamma_{S}}\right),
\end{align}
and the PDF of $\Gamma_{SE}$ can be expressed after differentiating (\ref{CDF_SNR_E}) as
\begin{align}\label{PDF_SNR_E}
&f_{{SE}}(x)=\frac{\frac{\lambda_{te}}{\Gamma_{T}}\exp\left(\frac{-\lambda_{se}x}{\Gamma_{S}}\right)}{x+\frac{\lambda_{te}\Gamma_{S}}{\lambda_{se}\Gamma_{T}}}+\frac{\frac{\lambda_{te}\Gamma_{S}}{\lambda_{se}\Gamma_{T}}\exp\left(\frac{-\lambda_{se}x}{\Gamma_{S}}\right)}{\left(x+\frac{\lambda_{te}\Gamma_{S}}{\lambda_{se}\Gamma_{T}}\right)^2}.
\end{align}

\section{Secrecy Outage Probability} \label{section 3}
In this section, we investigate the SOP of the secondary network. Towards deriving those performances, the secrecy capacity is required to be defined first. The secrecy capacity can be expressed for as
\begin{align}
C_{S} =\left[\log_2 (1+ \Gamma_{SD})-\log_2 (1+ \Gamma_{E})\right]^+ ,
\end{align}
where $\log_2 (1+ \Gamma_{SD})$ is the instantaneous capacity at D, $\log_2 (1+ \Gamma_E)$ is the instantaneous capacity of the wiretap channel at E, $\Gamma_{E}=\Gamma_{SE}$ and $[x]^{+}=\max(x,0)$. The SOP is defined as the probability that the secrecy rate is lower than a certain threshold, $R_{th}$. The SOP is given as
\begin{align}\label{secrecy OP equation}
\mathcal{P}_{out} (R_{th})
&= Pr(C_{S} < R_{th})\nonumber \\
&=\mathbb{P}\left[\Gamma_{SD}<R_{th}(1+\Gamma_{E})-1\right]
\nonumber \\
&= \int_{0}^{\infty} F_{SD}(\rho(x+1)-1) f_{SE}(x) dx,
\end{align}
where $\rho = 2^{R_{th}}-1$, $R_{th}$ is the target rate of the secondary network.
Substituting (\ref{OTS_CDF_SNR_TD}) and (\ref{PDF_SNR_E}) into (\ref{secrecy OP equation}), outage secrecy probability can be evaluated as
\begin{align}\label{NZ_rate OTS}
\mathcal{P}_{out} (R_{th})=1-A\cdot I_1-B\cdot I_2,
\end{align}
where $I_1$ and $I_2$ can be expressed respectively as 
\begin{align}\label{NZ I_1}
I_1=&\int_{0}^{\infty}\frac{1}{(x+a)(x+b)}\exp\left( -cx\right)dx,
\end{align}
\begin{align}\label{NZ I_2}
I_2=&\int_{0}^{\infty}\frac{1}{\left(x+a\right)\left(x+b\right)^2}\exp\left( -cx\right)dx,
\end{align}
with $a=\frac{\lambda_{td}\Gamma_{S}+k\rho\lambda_{sd}\Gamma_{T}-k\lambda_{sd}\Gamma_{T}}{k\rho\lambda_{sd}\Gamma_{T}}$, $b=\frac{\lambda_{te}\Gamma_S}{\lambda_{se}\Gamma_T}$ and $c=\frac{k\rho\lambda_{sd}+\lambda_{se}}{\Gamma_S}$
\begin{align}
A=&\sum_{k=1}^{K}\binom{K}{k}\Lambda(-1)^{k+1}\frac{\lambda_{te}\lambda_{td}\Gamma_{S}}{k\rho\lambda_{sd}\Gamma_{T}^2}\exp\left( \frac{-k\lambda_{sd}\left(\rho-1\right)}{\Gamma_{S}}\right),
\\
B=&\sum_{k=1}^{K}\binom{K}{k}\Lambda(-1)^{k+1}\frac{\lambda_{te}\lambda_{td}\Gamma_{S}^2}{k\rho\lambda_{se}\lambda_{sd}\Gamma_{T}^2}\exp\left( \frac{-k\lambda_{sd}\left(\rho-1\right)}{\Gamma_{S}}\right).
\end{align} 
We can utilize the partial fraction to transform multiplication into summation and solve (\ref{NZ I_1}) and (\ref{NZ I_2}) using 
\begin{align}\label{PF_NZ_OTS_I_1}
\frac{1}{(x+a)(x+b)}=-\frac{1}{(a-b)(x+a)}+\frac{1}{(a-b)(x+b)},
\end{align}
\begin{align}\label{PF_NZ_OTS_I_2}
\frac{1}{(x+a)(x+b)^2}&=\frac{1}{(a-b)^2(x+a)}-\frac{1}{(a-b)^2(x+b)}+\frac{1}{(a-b)(x+b)^2}.
\end{align}
For the final solution, we have used the integral solution of the form \cite{jeffrey2007table}, eq.(3.352.4) and \cite{jeffrey2007table}, eq.(3.353.3) to get
\begin{align}
I_1=&\frac{1}{a-b}\exp\left(ac\right)\mathrm{Ei}\left(-ac\right)-\frac{1}{a-b}\exp\left(bc\right)\mathrm{Ei}\left(-bc\right),
\\
I_2=&-\frac{1}{\left(a-b\right)^2}\exp\left(ac\right)\mathrm{Ei}\left(-ac\right)
+\frac{1}{\left(a-b\right)^2}\exp\left(bc\right)\mathrm{Ei}\left(-bc\right)
\nonumber \\
+&\frac{1}{a-b}\left(c\exp\left(bc\right)\mathrm{Ei}\left(-bc\right)+\frac{1}{b}\right).
\end{align}
\section{Numerical Results and Discussions}\label{section 5}
\begin{figure}
\centering
\includegraphics[trim=2cm 4cm 0.5cm 0cm, height=6cm, width=8cm]{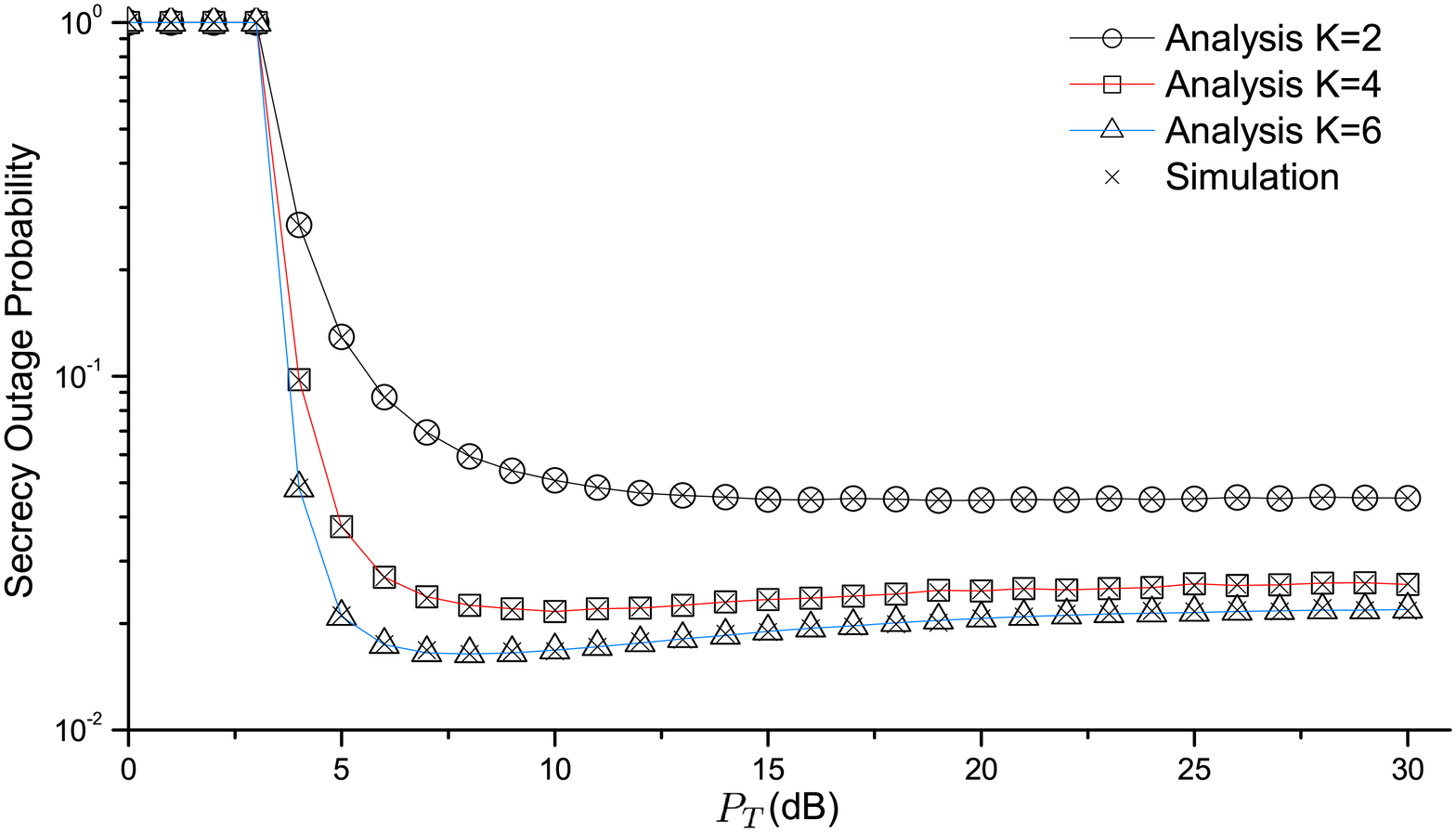} 
\caption{
SOP versus $P_T$(dB) for different numbers of secondary users.}
\label{fig:SOPvaryK}
\end{figure}

In this section, Monte Carlo simulations are provide to validate the theoretical analyses. Without loss of generality, we assume all nodes are affected by the same noise power $N_0$, and the following parameters are set: $\beta$=0.5 bits/s/Hz, $R_{th}=0.5$ bits/s/Hz, $\lambda_{tr}=3, \lambda_{td}=-6, \lambda_{sd}=3, \lambda_{sr}=-3, \lambda_{te}=6, \lambda_{se}=-3$ dB.

Fig. $\ref{fig:SOPvaryK}$ shows the SOP versus $P_T$ for different number of small-cell transmitters, K=2, K=4, and K=6. The network parameters are set as $\Phi=0.1$ and $\Lambda=0.99$. It shows that the analysis match with simulation. It can be observed that the number of small-cell transmitters strongly affects the SOP. As the number of smell-cell transmitter increase, SOP improves. However, the increase of transmitter i.e. K=2 to K=4 has more improvement compared to k=4 to K=6. As we increase $P_T$ SOP decreases first and converges to its floor  after certain values.

\begin{figure}
\centering
\includegraphics[trim=2cm 4cm 0.5cm 0cm, height=6cm, width=8cm]{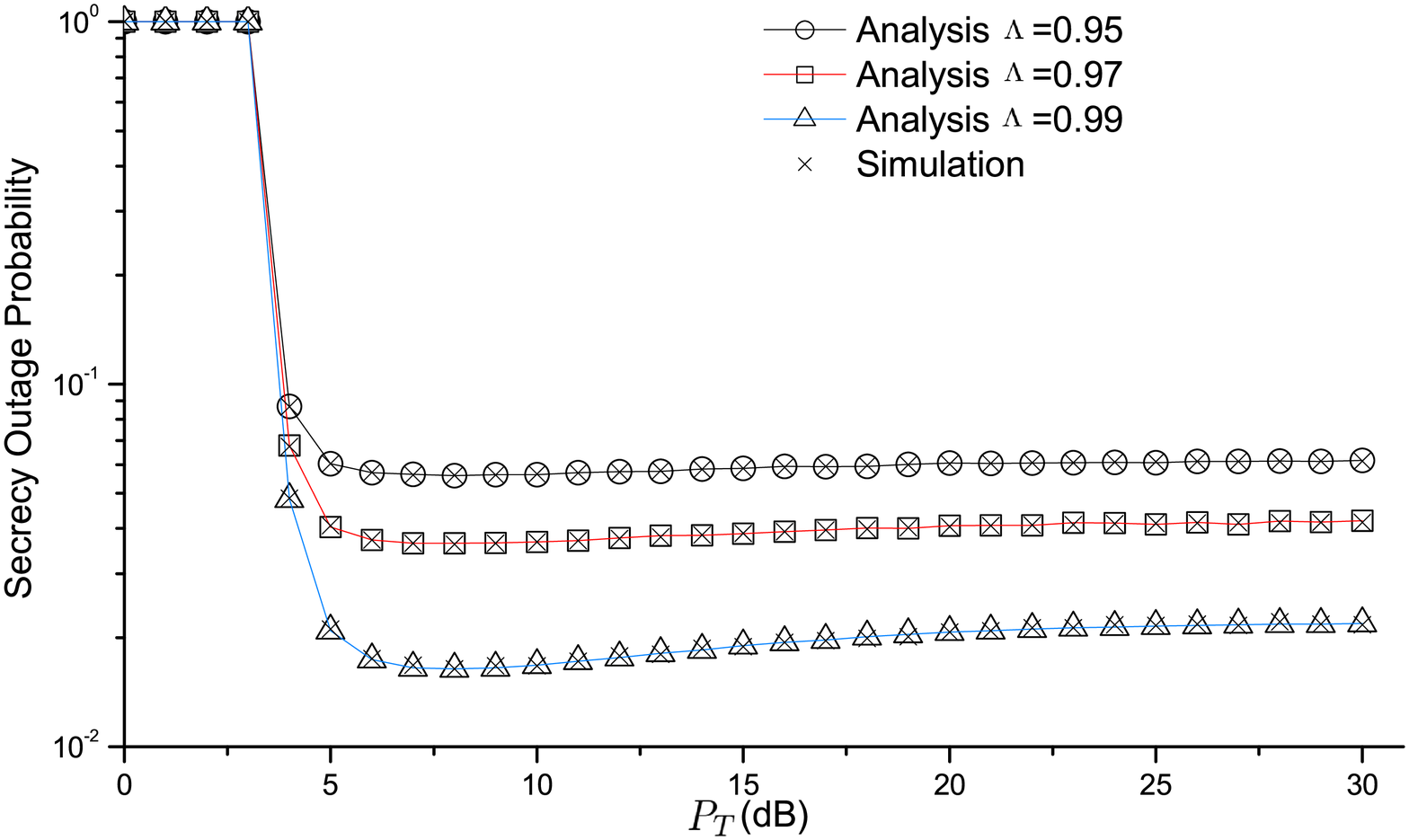} 
\caption{
SOP versus $P_T$(dB) for different values of $\Lambda$ .}
\label{fig:SOPvaryS}
\end{figure}

Fig. $\ref{fig:SOPvaryS}$ plots the SOP versus $P_T$ for different value of backhaul reliability, $\Lambda=0.95$ , $\Lambda=0.97$ and $\Lambda=0.99$ with K=6, $\Phi=0.1$. We observed that the SOP reduces when the $\Lambda$ increases. This is intuitive that as the reliability of the backhaul link improve of secrecy also improves. 

\begin{figure}
\centering
\includegraphics[trim=2cm 4cm 0.5cm 0cm, height=6cm, width=8cm]{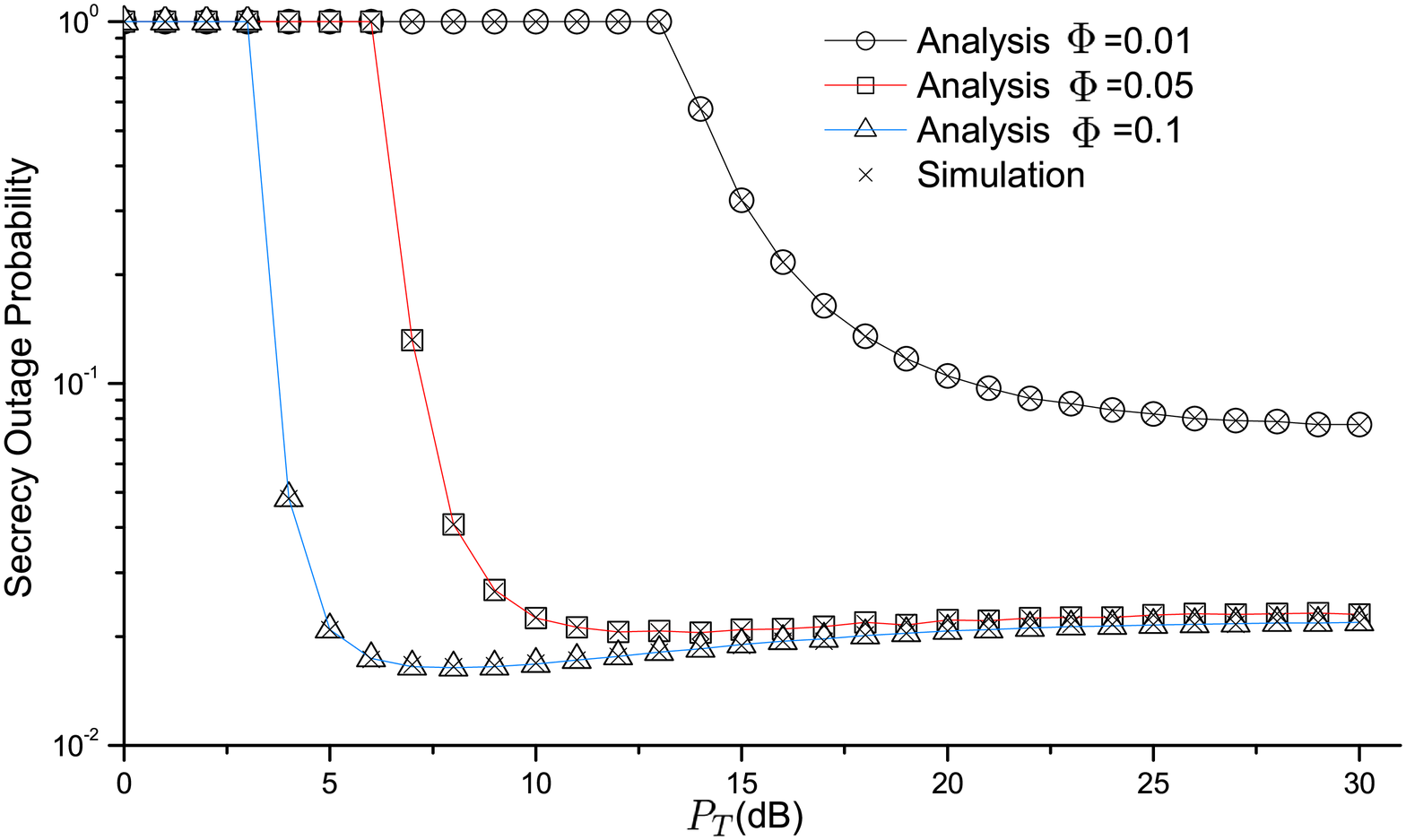} 
\caption{
SOP versus $P_T$(dB) for different values of $\Phi$.}
\label{fig:SOPQoS}
\end{figure}

In Fig. $\ref{fig:SOPQoS}$, the SOP is investigated versus $P_T$ with three different value of primary QoS constraint, $\Phi=0.01$, $\Phi=0.05$ and $\Phi=0.1$ with K=6, $\Lambda=0.99$. We observed that increasing $\Phi$ result in a reduction in the SOP. This is because the secondary network are allowed to have higher transmit power by relaxing the QoS requirement of the primary network. 
\section{Conclusion}\label{section 6}
In this paper, we have taken into account the backhaul connection reliability of studying the SOP of underlay cognitive radio network. A proposed selection scheme enhance the system's secrecy performance. The small-cell transmitter power met the desired outage probability. The results have proved that increasing the primary transmitter's power and the number of small-cell transmitter enhance the system's secrecy performance. In addition, our results shows that the backhaul reliability and the desired outage probability of the primary network are important parameter relative to the scaling of the secrecy performance. Increasing backhaul reliability will result in base station having higher success rate to connect with small-cell transmitters, and relaxing the QoS requirement of the primary network will allow small-cell transmitter to have higher transmit power, which will improve over all secrecy of the system.

%
%
%

\begin{thebibliography}{8}
\bibitem{Jeffrey20145G}
Andrews, J. G., Buzzi, S., Choi, W., \textit{et al}.: What will 5G be? \textit{IEEE Trans. Signal Process.} \textbf{32}(6), 1065--1082 (2014)

\bibitem{Orawan2011Evolution}
Tipmongkolsilp, O., Zaghloul, S., Jukan, A.: The evolution of cellular
backhaul technologies: Current issues and future trends. \textit{IEEE Commun. Surveys Tuts.} \textbf{13}(1), 97--113 (2011)

\bibitem{Xiaohu20145G}
Ge, X. H., Cheng, H., Guizani, M. \textit{et al}.: 5G wireless backhaul networks: challenges and research advances. \textit{IEEE Netw.} \textbf{28}(6), 6--11 (2014)

\bibitem{Kim2016Secrecy}
Kim, K. J., Yeoh, P. L., Orlik, P. V. \textit{et al}.: Secrecy Performance of Finite-Sized Cooperative Single Carrier Systems With Unreliable Backhaul Connections. \textit{IEEE Trans. Signal Process.} \textbf{64}(17), 4403--4416 (2016)

\bibitem{Kolodzy2002spectrum}
Kolodzy, P., Avoidance, Interference.: Spectrum policy task force. \textit{Federal Commun. Comm., Washington, DC, Rep. ET Docket} \textbf{40}(4), 147--158 (2002)

\bibitem{Mitola1999Cognitive}
Mitola, J.,  Maguire, G. Q.: Cognitive radio: making software radios more personal. \textit{IEEE Personal Commun.} \textbf{6}(4), 13-18 (1999)

\bibitem{Zhangjinghua}
Zhang, J. H., Nguyen, N. P., Zhang, J. Q., \textit{et al}.: Impact of primary networks on the performance of energy harvesting cognitive radio networks. \textit{IET Commun.} \textbf{10}(18), 2559-2566 (2016)

\bibitem{Lee2015Cognitive}
Lee, S., Zhang, R.: Cognitive wireless powered network: Spectrum sharing models and throughput maximization. \textit{IEEE Trans. Cognitive Commun.} \textbf{1}(3), 335--346 (2015)

\bibitem{Phong2016Secure}
Nguyen, N. P., Duong, T. Q., Ngo, H. Q., \textit{et al}.: Secure 5G Wireless Communications: A Joint Relay Selection and Wireless Power Transfer Approach. \textit{IEEE Access} \textbf{4}(), 3349--3359 (2016)

\bibitem{Trung2013Cognitive}
Bao, V. N. Q., Duong, T. Q., Da Costa, D. B., \textit{et al}.: Cognitive amplify-and-forward relaying with best relay selection in
non-identical Rayleigh fading. \textit{IEEE Commun. Lett.} \textbf{17}(3), 475--478 (2013)

\bibitem{Kundu2016Relay}
Kundu, C., Ngatched, T. M. N., Dobre, O. A.: Relay selection to improve secrecy in cooperative threshold decode-and-forward relaying. \textit{in Proc. IEEE GLOBECOM 2016}, Washington, DC, USA, 4--8 (2016)

\bibitem{jinghuazhang2017Cognitive}
Zhang, J., Kundu, C., Nguyen, N. P., \textit{et al}.: Cognitive Wireless Powered Communication Networks with Secondary User Selection and Primary QoS Constraint. \textit{IET Commun.} in press

\bibitem{Yuzhen2016Secure}
Huang, Y. Z., Wang, J. L., Zhong, C. J., \textit{et al}.: Secure transmission in cooperative relaying networks with multiple antennas. \textit{IEEE Trans. Wireless Commun.} \textbf{15}(10), 6843--6856 (2016)

\bibitem{Yincheng}
Yin, C., Nguyen, H. T., Kundu, C., \textit{et al}.: Secure energy harvesting relay networks with unreliable backhaul connections. \textit{IEEE Access} \textbf{6}(), 12074--12084 (2018)

\bibitem{Vu2017Secure}
Vu, T., Nguyen, M. N., Kundu, C., \textit{et al}.: Secure cognitive radio networks with source selection and unreliable backhaul connections. \textit{IET Commun.} \textbf{}(), (2017)

\bibitem{Kundu2017AFrelay}
Kundu, C., Jindal, A., Bose, R.: Secrecy outage of dual-hop amplify-and-forward relay system with diversity combining at the eavesdropper. \textit{Wireless Personal Commun.} \textbf{}(), (2017)

\bibitem{Kim2015Performance}
Khan, T. A., Orlik, P., Kim, K. J. \textit{et al}.: Performance analysis of cooperative
wireless networks with unreliable backhaul links. \textit{IEEE Commun. Lett.} \textbf{19}(8), 1386--1389 (2015)

\bibitem{Huy2017Multiuser}
Nguyen, H. T., Duong, T. Q., Hwang, W. J.: Multiuser relay networks over unreliable backhaul links under spectrum sharing environment. \textit{IEEE Commun. Lett.} \textbf{21}(10), 2314--2317 (2017)

\bibitem{Huy2017Secure}
Nguyen, H. T., Zhang, J. Q., Yang, N., \textit{et al}.: Secure Cooperative Single Carrier Systems Under Unreliable Backhaul and Dense Networks Impact. \textit{IEEE Access} \textbf{5}(), 18310--18324 (2017)

\bibitem{jeffrey2007table}
Jeffrey, A., Zwillinger, D.: Table of integrals, series, and products. 7nd edn. Academic press, (2007)
\end{thebibliography}
\end{document}